\documentclass[%
 reprint,
 amsmath,amssymb,
 aps,
]{revtex4-1}

\usepackage{graphicx}
\usepackage{dcolumn}
\usepackage{bm}

\begin{document}

\preprint{APS/123-QED}

\title{Thermal properties of ‘athermal’ granular materials}

\author{Kasra Farain and Daniel Bonn}

\affiliation{Van der Waals–Zeeman Institute, Institute of Physics, University of Amsterdam, Science Park 904,  1098XH Amsterdam, The Netherlands\\
 k.farain@uva.nl\\d.bonn@uva.nl}

\begin{abstract}
Dry granular materials consist of a vast ensemble of discrete solid particles, interacting through complex frictional forces at the contact points. The particles are so large that these systems are believed to be completely athermal. Here, we arrest the dynamics of a flowing granular material in a steady-state flow configuration, enabling an isolated examination of aging at the particle contacts without granular rearrangements. Our findings reveal that the evolution of interparticle forces within the arrested athermal granular network results in the spontaneous increase of the system’s yield stress. This strengthening process is logarithmic in time with a rate that depends on temperature. We demonstrate that the material’s stress relaxation exhibits similar time- and temperature-dependent behavior, suggesting a shared origin for aging and stress relaxation in these systems governed by thermal molecular processes at the scale of the grain contacts.
\end{abstract}

\pacs{Valid PACS appear here}
\maketitle



The mechanics of dry granular materials, characterized by large particles lacking thermal motion, hold important practical relevance across various industries, from processing plastics in granular form to construction projects involving sand mounds. Beyond industrial applications, these materials play a central role in natural events such as avalanches, landslides, and earthquakes, where the behavior of fault gouge—a granular substance within seismic fault zones—dictates fault dynamics \cite{1, 2, 3, 4}. Granular materials possess a dual nature: in many cases, they can support a load like a solid, yet under large external forces, they can flow akin to a liquid \cite{5, 6}. The transition from solid-like packing to liquid-like flow occurs at a critical threshold stress, often referred to as the material's yield stress. However, this threshold stress is not an inherent material property; rather, it can vary based on experimental preparations and conditions and the material history \cite{1, 7}. In particular, the shear stress required to initiate flow may demonstrate a strengthening or aging effect over time under specific conditions \cite{8, 9}.

The strengthening effect within a granular packing can arise from the evolution of the spatial arrangement of particles, i.e. the granular network or configuration, or the amplification of interparticle forces. For instance, in Bocquet et al.'s experiments \cite{8} on a system of glass beads within a rotating drum, a strengthening effect was observed due to the formation of liquid bridges, introducing an attractive capillary force between the beads. This effect was characterized by a logarithmic increase in the angle of avalanche of the granular material, displaying a rate of increase tied to humidity level, with no discernible effect observed at low humidity levels. In contrast, Losert et al's experiments \cite{9} revealed a roughly logarithmic increase in the strength of packings of glass beads both underwater and in a dry environment, where the absence of liquid bridges and capillary forces was evident. However, this strengthening exclusively occurred when a shear stress was applied during the waiting period. Their work proposed that the strengthening of granular material stems from gradual changes in particle arrangement under shear stress. Similarly, studies conducted under geophysical pressures (20 MPa) on granular quartz powder reported a logarithmic strengthening of granular layers, but again, specifically when a shear stress was applied during the waiting period \cite{3, 4}.

In this Letter, we show that there may be another origin for a time-evolution of granular systems. We demonstrate that in a flowing granular system, a slight reduction in the applied shear stress can cause the system to fall below its dynamic yield stress, leading to the freezing of the flow configuration: Essentially, the flow ceases while the spatial arrangement of beads remains in the steady-state flow configuration. By freezing a granular material at steady-state flow and reinstating flow directly from this state without configurational-related transients, we isolate the effect of aging of the interparticle contacts. This approach reveals a spontaneous development of a friction peak within the granular packing that grows logarithmically with the hold time. This growth is linked to the strengthening of discrete microcontacts between grains due to slow creep deformations. Our measurements indicate an amplification of this aging effect at higher temperatures, revealing the thermal nature of aging in such athermal systems. Furthermore, the stress relaxation rate in the material across different temperatures is similar, hinting at a shared origin for aging and stress relaxation through thermal processes within these systems.

\begin{figure}
	\begin{center}
		\includegraphics[scale=1]{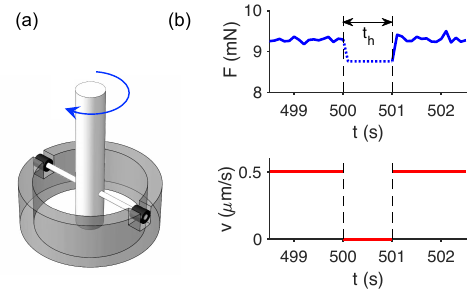}
	\end{center}
	\vspace{-0.5cm}
	\caption{Arresting the granular flow configuration. (a) Schematic representation of the custom-made rheology measuring system used in the study. (b) A constant sliding rate of $v=0.5 ~\mu$m/s ($10^{-5}~s^{-1}$ rotational speed in the rheometer) is applied to the system until it reaches a SS flow condition. At time $t_{1} = 500$ s, the constant shear rate is switched to a constant force (torque) slightly smaller than the SS friction of the granular system, leading to flow cessation, $v=0$. Then, at $t_{2} = 501$ s, the granular material is again subjected to  $v=0.5 ~\mu$m/s to re-initiate flow. However, since the material maintained its granular flow arrangement during the stop phase, the transient dynamics normally observed at the onset of granular flows are absent. Conversely, any disruption of the shear stress causes the collapse of this flow configuration, leading to the resurgence of configuration-related transients upon reintroduction of flow to the material.}
	\label{f:numeric}\end{figure}

\begin{figure*}[hbt]
	\begin{center}
		\includegraphics[scale=1]{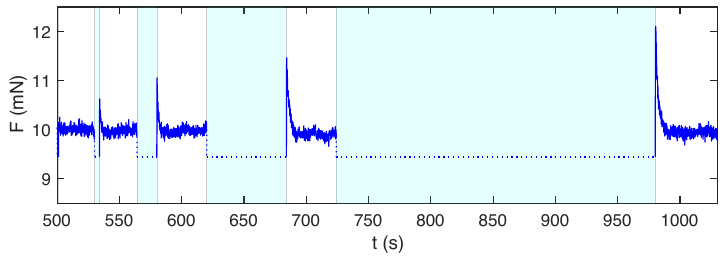}
	\end{center}
	\vspace{-.5cm}
	\caption{Frictional aging at a frozen granular flow configuration. The stop-and-start experiments, described in Fig. 1, were conducted with longer hold times (blue shading). As the hold time is increased, a growing friction peak is observed at the time the system is made to flow again.}
	\label{f:numeric}\end{figure*}

Our experimental system \cite{7}, schematically shown in Fig. 1(a), is comprised of an aluminum ring (outer diameter of $29.7$ mm, weight of 65 mN) with a thick edge ($2.3$ mm) positioned on a bed of granular material composed of hard, spherical poly(methylmethacrylate) (PMMA) micro-particles with an approximate diameter of  $40~\mu$m. The ring is connected to an Anton Paar MCR 302 rheometer via a custom-made measuring system, enabling its free movement in vertical direction and settling on the granular material. The granular material was spread evenly onto an aluminum plate under the rheometer head. The ring is then slowly (speed of a few $\mu$m/s) positioned on top of the granular layer, ensuring proper alignment with the arms of the rheometer measuring system. The measuring system is subsequently lowered into predefined slots in the ring. Low-friction bearings were incorporated between the rheometer tool and the ring, which guarantees that the granular material experienced a constant normal load corresponding to the weight of the ring throughout the experiments. To prevent wall slip, both the ring and bottom plate frictional surfaces were sandblasted to reach a surface roughness of approximately $2–3 ~\mu$m. This was further verified by observing consistent results even after applying a layer of adhesive material to the aluminum frictional surfaces. To heat the granular material during the experiments, the rheometer's temperature plate (Peltier Temperature Device) was employed.

The rheometer possesses the capability to apply remarkably low rotational speeds, reaching as low as $~ 10^{-8} ~s^{-1}$. This precision allows for the precise and uniform movement of the macroscopic ring with nanometer resolution, a crucial aspect in arresting granular flow at a steady-state configuration without causing disruptions. Moreover, the rheometer enables direct switching between different modes of motion without releasing stresses on the system in between, which is essential for our experiments. Specifically, it can seamlessly transition between applying a rotational torque to the ring while measuring the resulting rotational speed and applying the rotational speed while measuring the torque.

We implement an experimental procedure in which the granular flow configuration is frozen in time, and after a waiting time, continued again without any change in the material. This is possible due to the first-order nature of yield stress material: By carefully adjusting the applied shear stress to be just below or above the (dynamic) yield stress $F_{SS}$, we can pause or initiate the flow while keeping the stresses within the material, and thus its configuration, largely unchanged. In a typical experiment (see Fig. 1(b)), we allow the granular material to reach a SS flow configuration under a specific shear rate. At time $t_{1}=500$ s, the rheometer switches to a constant shear stress slightly below the force required to continue the SS flow $F_{SS}$, leading to the cessation of the granular flow. However, the material should maintain its SS configuration as the stresses within the material haven’t changed substantially: The same ensemble of microforces in the granular network is in equilibrium with the external shear and normal loads on the system. Following a 1 s pause, at time $t_{2}=501$ s, we revert to the constant shear rate. Remarkably, despite the interruption and resumption of flow, we observe no transient stress dynamics typically seen during the initiation of flow in these systems. The shear stress response (friction) restarts precisely from where it left off.

The notion that the granular configuration remains unchanged when fine-tuning shear stress slightly below or above the yield stress can be interpreted from different viewpoints. A granular material model, considering stress chains among grains, illustrates how these chains dynamically adapt and redistribute under external stresses \cite{10, 11}. As long as the external forces acting upon the granular material's boundary remain constant, there's no necessity for new force-chain rearrangements to sustain equilibrium. The observed stability of configuration during the above cessation and restart of flow can also be comprehended within the context of the quasi-stationary nature of granular flow in our experiments, where inertia is negligible. The absence of inertia to be dissipated by pausing the flow, or any other active mechanisms for elastic energy dissipation in the static state (except the slow stress relaxation that will be discussed below), indicates that this stop-and-start method can be compared to pausing and resuming a video of flowing granular material. When pausing the 'video' or abruptly locking the rheometer's rotational motion, all granular-level dynamics within the system freeze instantly at their current state.

Contrarily, when the shear stress is completely or partially removed from the system during the stop phase, the flow configuration collapses under the ring's weight. This collapse occurs as the fragile network of stress chains breaks apart, giving rise to new force chains that align with the altered external forces. Subsequently, upon reintroducing the shear rate in this case, a continuous prolonged stress overshoot emerges, signifying a reconfiguration within the material \cite{7}. This depiction concurs with prior observations that emphasize a transient phase within the material when the shear direction is reversed \cite{12}.

While on a macroscopic scale, grains are regarded as constituent elements of the material which determine its overall rheological properties, at a microscopic level, these grains can be perceived as stationary solid entities in frictional contact. Especially, the particles are large enough to display no Brownian dynamics, and the system experiences no external vibrations (the experiments are conducted on an optical table). Consequently, under small forces below a specific threshold, the pile of discrete particles reaches a state of mechanical equilibrium determined by the external forces acting on the system. In the above stop-and-start experiments, this mechanical equilibrium is the frozen SS flow configuration.

In solid-on-solid friction, it is known that the frictional threshold keeping two surfaces together typically undergoes a logarithmic increase with the duration of stationary contact before sliding. This phenomenon has been observed across a wide spectrum of materials and systems, ranging from paper \cite{13} to granite rocks \cite{14}, and in both macroscopic and nanoscale frictional systems \cite{15, 16}. Within non-sheared granular material with a completely random contact networks, observing frictional aging may be improbable: The onset of flow involves individual inter-particle contact points breaking up separately, allowing particles to gradually rearrange in response to the imposed shear. Thus, the granular system's stress response primarily reflects this rearrangement, resulting in the continuous prolonged stress overshoot \cite{7}. In contrast, upon reapplying shear rate to a frozen SS-flow configuration, no stress changes due to granular rearrangements are observed as flow promptly resumes from the correct configuration.  The question is whether frictional aging between grains manifests under this condition. Even within SS flow, the sheared regions in granular materials exhibit complex three-dimensional characteristics, wherein neighboring particles display varying directions and velocities \cite{9}. Moreover, not all contact points experience shear motion; particles might roll or detach, involving adhesion forces perpendicular to the contact interface. Consequently, the laws governing solid friction, associated with a two-dimensional array of asperity contacts slipping, might not directly apply to 3D granular systems. However, frictional aging might merely represent one facet of the logarithmic evolution of contacts between solids, influenced by factors such as creep and other alterations leading to a redistribution of contact stresses. Despite the diversity in forces and their directions, the evolution of inter-particle contacts may still lead to the collective strengthening of the granular system.

To explore interparticle aging within the granular packing, we perform a series of stop-and-start experiments with extended hold times (Fig. 2). When subjecting a granular system, frozen in its SS configuration for a few seconds or longer, to a constant shear rate, we observe a distinct spike in friction before reaching the constant SS value. These new stress peaks are wholly different entities than the continuous stress overshoots observed at the onset of granular flows involving configurational changes \cite{7}. The newly observed friction peaks exhibit a rapid linear ascent to a maximum value, followed by an essentially discontinuous failure. In Fig. 3, the friction spike height normalized by SS friction is plotted as a function of the hold time at various temperatures. Consistent with the dynamics of general frictional aging between solids, the normalized height of these friction spikes increases logarithmically with hold time $t_{h}$. We can write

\begin{equation}
	\frac{F_{Peak}}{F_{SS}}=\Delta_{T}\ln(\frac{t_{h}}{1 s})+C,
\end{equation}
where $\Delta_{T}$ is the temperature-dependent aging rate (a small dimenstionless number), and $C$ is a constant close to 1, both obtained from the experiments. It's worth noting that the scope of Eq. 1 is limited to $t_{h}\gtrsim 1$ s, below which $F_{Peak} \approx F_{SS}$. The logarithmic evolution continues for at least three orders of magnitude in time.

Figure 3, additionally, depicts a notable increase in the development rate of the frictional spike at higher temperatures. This clearly signals the thermally-driven molecular nature of such strengthening. At this scale, the temperature-dependent molecular chain mobility of polymers governs the material's slow relaxation and aging \cite{17, 18, 19}. In fact, previous research has linked frictional aging to bulk stress relaxation in materials \cite{20}: Macroscopic spheres of polypropylene or polytetrafluoroethylene (PTFE) on glass substrates exhibited logarithmic frictional aging, with dimensionless aging rates of 0.042 and 0.036, respectively, at room temperature. These same rates were obtained from independent bulk stress relaxation experiments \cite{20}. Interestingly, these values for polypropylene and PTFE closely resemble the aging rate $\Delta_{T}=0.033$ found in our current study for PMMA granular material at 20 $^{\circ}$C. 

\begin{figure}[hbt]
	\begin{center}
		\includegraphics[scale=1]{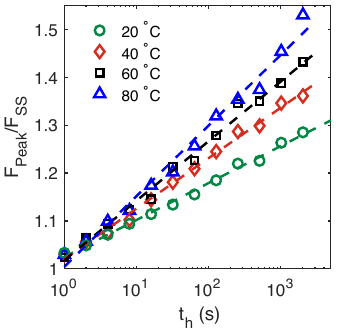}
	\end{center}
	\vspace{-.5cm}
	\caption{Frictional aging in granular materials at varying temperatures. Friction spike height from the stop-and-start experiments, normalized by the SS friction as a function of the aging (hold) time at varying temperatures. The dashed lines represent the best fits of Eq. 1, yielding $\Delta_{T}$ values of 0.033, 0.046, 0.054, and 0.064 at temperatures of 20, 40, 60, and 80 $^{\circ}$C, respectively.}
	\label{f:numeric}\end{figure}

\begin{figure}[hbt]
	\begin{center}
		\includegraphics[scale=1]{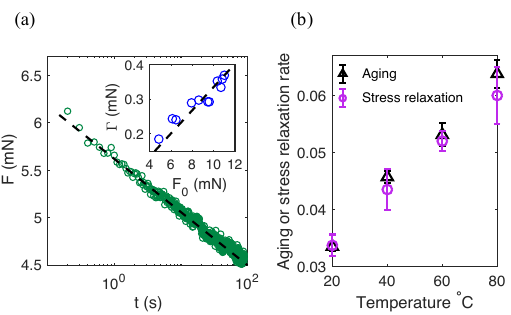}
	\end{center}
	\vspace{-.5cm}
	\caption{Stress relaxation and aging have similar temperature-dependence. (a) Shear force applied to a sample of the granular material to keep a constant deformation at 20 $^{\circ}$C. This force exhibits a slow logarithmic relaxation over time. The inset displays the slope of the logarithmic relaxation $\Gamma$, as defined in Eq. (2), derived from multiple experiments conducted at this temperature as a function of the initial applied shear stress, $F_{0}$. The dashed line $\Gamma=\alpha F_{0}$ represents a linear fit passing through the origin, reflecting the fact that zero shear force corresponds to zero relaxation. (b) Dimensionless aging ($\Delta_{T}$) and stress relaxation ($\alpha$) rates observed at various temperatures. The graph illustrates the similar temperature-dependence of these two phenomena.}
	\label{f:numeric}\end{figure}

We now turn our attention to exploring the stress relaxation characteristic of the granular material as a whole. In these experiments, the rheometer applies a predefined shear deformation to the material, resulting in the buildup of a shear stress. Then it tracks how that stress changes over time while keeping the deformation constant. In Fig. 4(a), the result from a typical experiment at 20 $^{\circ}$C is displayed. We note a logarithmic pattern in stress relaxation, similar to the characteristic behavior observed in frictional aging. This can be described by
\begin{equation}
F=\Gamma \ln(\frac{t}{\tau}),
\end{equation}
where $F$ is shear force, $\Gamma$  is an experiment-specific constant, and $\tau$ is a time scale. The inset in Fig. 4(a) shows how $\Gamma$, the logarithmic behavior's slope, relates to the initial applied stress. These results are fitted with a linear trend that intersects at the origin ($\Gamma=\alpha F_{0}$), as zero shear force must denote no relaxation and, therefore, $\Gamma=0$. Figure 4(b) combines $\alpha$ values from more stress relaxation experiments at different temperatures with aging rates from Fig. 3. These results highlight a clear temperature-dependent relationship between frictional aging and stress relaxation. This connection suggests that the processes causing these phenomena at a molecular level might be identical. Indeed, in both aging within a frozen configuration and static stress relaxation experiments, the athermal granular configuration remains relatively stable. However, the molecular configuration of the system may undergo evolution due to thermally-driven effects, and it's this evolution that we measure in these different experiments.

Overall, our research delineates the influence of thermally driven molecular processes and athermal configurational changes on the stability of granular systems, each giving rise to a distinct yield stress within the material. In non-sheared granular packing, the yield stress arises from the stress overshoot due to structural changes under shear. If an applied stress, $\Sigma$, exceeds the peak stress overshoot, $F_{Peak}$, the material starts flowing; otherwise, the material remains stable. $F_{Peak}$ varies irregularly across different preparations of the system, depending on the packing's arrangement \cite{7}. Conversely, thermal aging introduces another type of yield stress linked to inter-particle contact aging in previously sheared systems, rising over time at a defined temperature-dependent logarithmic rate. Such scenarios manifest in phenomena like stick-slip dynamics, wherein the system pauses at the flow configuration, and in conical granular piles where the outer layers approach the flow threshold.

We emphasize the central role played by thermal molecular-level processes in the evolution of granular material. Traditionally, the observed strengthening behavior in granular materials and dense suspensions over time has been predominantly attributed to structural changes and compaction within these systems. Through the arrest of granular flow configurations in SS, we have effectively isolated the inherent molecular-level changes present in these systems at any non-zero temperature. Our findings suggest that the same thermal molecular-level processes, responsible for strengthening granular materials under shear stress, also dictate stress relaxation under a constant applied deformation. Future investigations should develop molecular models that align with observed temperature dependence of stress relaxation and aging in granular materials.

Furthermore, we have noted intriguing similarities between aging phenomena in dry granular material and those observed in solid-on-solid friction, despite their distinct geometries. Considering dry granular material as a model system for an ensemble of frictional asperity contacts presents a promising approach to overcome reproducibility challenges in friction experiments, allowing for deeper insights into the dynamics of solid-on-solid friction in general. 

\vspace*{0.3 cm}
This project has received funding from the European Research Council (ERC) under the European Union’s Horizon 2020 research and innovation program (Grant agreement No. 833240)

\end{document}